%
%
%
\def\unredoffs{} \def\redoffs{\voffset=-.31truein\hoffset=-.59truein}
\def\speclscape{\special{ps: landscape}}
%
%
%
%
\newbox\leftpage \newdimen\fullhsize \newdimen\hstitle \newdimen\hsbody
\tolerance=1000\hfuzz=2pt
\catcode`\@=11 
\def\bigans{b }
\def\answ{b }
\ifx\answ\bigans\message{(This will come out unreduced.}
\magnification=1200\unredoffs\baselineskip=16pt plus 2pt minus 1pt
\hsbody=\hsize \hstitle=\hsize 
\else\message{(This will be reduced.} \let\l@r=L
\magnification=1000\baselineskip=16pt plus 2pt minus 1pt \vsize=7truein
\redoffs \hstitle=8truein\hsbody=4.75truein\fullhsize=10truein\hsize=\hsbody
\output={\ifnum\pageno=0 
  \shipout\vbox{\speclscape{\hsize\fullhsize\makeheadline}
    \hbox to \fullhsize{\hfill\pagebody\hfill}}\advancepageno
  \else
  \almostshipout{\leftline{\vbox{\pagebody\makefootline}}}\advancepageno
  \fi}
\def\almostshipout#1{\if L\l@r \count1=1 \message{[\the\count0.\the\count1]}
      \global\setbox\leftpage=#1 \global\let\l@r=R
 \else \count1=2
  \shipout\vbox{\speclscape{\hsize\fullhsize\makeheadline}
      \hbox to\fullhsize{\box\leftpage\hfil#1}}  \global\let\l@r=L\fi}
\fi
%
\newcount\yearltd\yearltd=\year\advance\yearltd by -1900

\def\Title#1#2{\nopagenumbers\abstractfont\hsize=\hstitle\rightline{#1}%
\vskip 1in\centerline{\titlefont #2}\abstractfont\vskip .5in\pageno=0}
\def\Date#1{\vfill\leftline{#1}\tenpoint\supereject\global\hsize=\hsbody%
\footline={\hss\tenrm\folio\hss}}
%

\def\draftmode{\message{ DRAFTMODE }\def\draftdate{{\rm preliminary draft:
\number\month/\number\day/\number\yearltd\ \ \hourmin}}%
\headline={\hfil\draftdate}\writelabels\baselineskip=20pt plus 2pt minus 2pt
 {\count255=\time\divide\count255 by 60 \xdef\hourmin{\number\count255}
  \multiply\count255 by-60\advance\count255 by\time
  \xdef\hourmin{\hourmin:\ifnum\count255<10 0\fi\the\count255}}}
\def\nolabels{\def\wrlabeL##1{}\def\eqlabeL##1{}\def\reflabeL##1{}}
\def\writelabels{\def\wrlabeL##1{\leavevmode\vadjust{\rlap{\smash%
{\line{{\escapechar=` \hfill\rlap{\sevenrm\hskip.03in\string##1}}}}}}}%
\def\eqlabeL##1{{\escapechar-1\rlap{\sevenrm\hskip.05in\string##1}}}%
\def\reflabeL##1{\noexpand\llap{\noexpand\sevenrm\string\string\string##1}}}
\nolabels
%
\global\newcount\secno \global\secno=0
\global\newcount\meqno \global\meqno=1
\def\newsec#1{\global\advance\secno by1\message{(\the\secno. #1)}
\global\subsecno=0\eqnres@t\noindent{\bf\the\secno. #1}
\writetoca{{\secsym} {#1}}\par\nobreak\medskip\nobreak}
\def\eqnres@t{\xdef\secsym{\the\secno.}\global\meqno=1\bigbreak\bigskip}
\def\sequentialequations{\def\eqnres@t{\bigbreak}}\xdef\secsym{}
\global\newcount\subsecno \global\subsecno=0
\def\subsec#1{\global\advance\subsecno by1\message{(\secsym\the\subsecno.
#1)}
\ifnum\lastpenalty>9000\else\bigbreak\fi
\noindent{\it\secsym\the\subsecno. #1}\writetoca{\string\quad
{\secsym\the\subsecno.} {#1}}\par\nobreak\medskip\nobreak}
\def\appendix#1#2{\global\meqno=1\global\subsecno=0\xdef\secsym{\hbox{#1.}}
\bigbreak\bigskip\noindent{\bf Appendix #1. #2}\message{(#1. #2)}
\writetoca{Appendix {#1.} {#2}}\par\nobreak\medskip\nobreak}
%
%
\def\eqnn#1{\xdef #1{(\secsym\the\meqno)}\writedef{#1\leftbracket#1}%
\global\advance\meqno by1\wrlabeL#1}
\def\eqna#1{\xdef #1##1{\hbox{$(\secsym\the\meqno##1)$}}
\writedef{#1\numbersign1\leftbracket#1{\numbersign1}}%
\global\advance\meqno by1\wrlabeL{#1$\{\}$}}
\def\eqn#1#2{\xdef #1{(\secsym\the\meqno)}\writedef{#1\leftbracket#1}%
\global\advance\meqno by1$$#2\eqno#1\eqlabeL#1$$}
%
\newskip\footskip\footskip14pt plus 1pt minus 1pt 
\def\footnotefont{\ninepoint}\def\f@t#1{\footnotefont #1\@foot}
\def\f@@t{\baselineskip\footskip\bgroup\footnotefont\aftergroup\@foot\let\next}
\setbox\strutbox=\hbox{\vrule height9.5pt depth4.5pt width0pt}
\global\newcount\ftno \global\ftno=0
\def\foot{\global\advance\ftno by1\footnote{$^{\the\ftno}$}}
%
\newwrite\ftfile
\def\footend{\def\foot{\global\advance\ftno by1\chardef\wfile=\ftfile
$^{\the\ftno}$\ifnum\ftno=1\immediate\openout\ftfile=foots.tmp\fi%
\immediate\write\ftfile{\noexpand\smallskip%
\noexpand\item{f\the\ftno:\ }\pctsign}\findarg}%
\def\footatend{\vfill\eject\immediate\closeout\ftfile{\parindent=20pt
\centerline{\bf Footnotes}\nobreak\bigskip\input foots.tmp }}}
\def\footatend{}
%
%
\global\newcount\refno \global\refno=1
\newwrite\rfile
\def\ref{[\the\refno]\nref}
\def\nref#1{\xdef#1{[\the\refno]}\writedef{#1\leftbracket#1}%
\ifnum\refno=1\immediate\openout\rfile=refs.tmp\fi
\global\advance\refno by1\chardef\wfile=\rfile\immediate
\write\rfile{\noexpand\item{#1\ }\reflabeL{#1\hskip.31in}\pctsign}\findarg}
\def\findarg#1#{\begingroup\obeylines\newlinechar=`\^^M\pass@rg}
{\obeylines\gdef\pass@rg#1{\writ@line\relax #1^^M\hbox{}^^M}%
\gdef\writ@line#1^^M{\expandafter\toks0\expandafter{\striprel@x #1}%
\edef\next{\the\toks0}\ifx\next\em@rk\let\next=\endgroup\else\ifx\next\empty%
\else\immediate\write\wfile{\the\toks0}\fi\let\next=\writ@line\fi\next\relax}}
\def\striprel@x#1{} \def\em@rk{\hbox{}}
\def\lref{\begingroup\obeylines\lr@f}
\def\lr@f#1#2{\gdef#1{\ref#1{#2}}\endgroup\unskip}
\def\semi{;\hfil\break}
\def\addref#1{\immediate\write\rfile{\noexpand\item{}#1}} 
\def\startrefs#1{\immediate\openout\rfile=refs.tmp\refno=#1}
\def\xref{\expandafter\xr@f}\def\xr@f[#1]{#1}
\def\refs#1{\count255=1[\r@fs #1{\hbox{}}]}
\def\r@fs#1{\ifx\und@fined#1\message{reflabel \string#1 is undefined.}%
\nref#1{need to supply reference \string#1.}\fi%
\vphantom{\hphantom{#1}}\edef\next{#1}\ifx\next\em@rk\def\next{}%
\else\ifx\next#1\ifodd\count255\relax\xref#1\count255=0\fi%
\else#1\count255=1\fi\let\next=\r@fs\fi\next}
%

%
\newwrite\ffile\global\newcount\figno \global\figno=1
\def\fig{fig.~\the\figno\nfig}
\def\nfig#1{\xdef#1{fig.~\the\figno}%
\writedef{#1\leftbracket fig.\noexpand~\the\figno}%
\ifnum\figno=1\immediate\openout\ffile=figs.tmp\fi\chardef\wfile=\ffile%
\immediate\write\ffile{\noexpand\medskip\noexpand\item{Fig.\ \the\figno. }
\reflabeL{#1\hskip.55in}\pctsign}\global\advance\figno by1\findarg}
\def\xfig{\expandafter\xf@g}\def\xf@g fig.\penalty\@M\ {}
\def\figs#1{figs.~\f@gs #1{\hbox{}}}
\def\f@gs#1{\edef\next{#1}\ifx\next\em@rk\def\next{}\else
\ifx\next#1\xfig #1\else#1\fi\let\next=\f@gs\fi\next}
\newwrite\lfile
{\escapechar-1\xdef\pctsign{\string\%}\xdef\leftbracket{\string\{}
\xdef\rightbracket{\string\}}\xdef\numbersign{\string\#}}

\def\writestop{\def\writestoppt{\immediate\write\lfile{\string\pageno%
\the\pageno\string\startrefs\leftbracket\the\refno\rightbracket%
\string\def\string\secsym\leftbracket\secsym\rightbracket%
\string\secno\the\secno\string\meqno\the\meqno}\immediate\closeout\lfile}}
\def\writestoppt{}\def\writedef#1{}
\def\seclab#1{\xdef #1{\the\secno}\writedef{#1\leftbracket#1}\wrlabeL{#1=#1}}
\def\subseclab#1{\xdef #1{\secsym\the\subsecno}%
\writedef{#1\leftbracket#1}\wrlabeL{#1=#1}}
\newwrite\tfile \def\writetoca#1{}
\def\leaderfill{\leaders\hbox to 1em{\hss.\hss}\hfill}
\def\writetoc{\immediate\openout\tfile=toc.tmp
   \def\writetoca##1{{\edef\next{\write\tfile{\noindent ##1
   \string\leaderfill {\noexpand\number\pageno} \par}}\next}}}

%
\catcode`\@=12 
%
\edef\tfontsize{\ifx\answ\bigans scaled\magstep3\else scaled\magstep4\fi}
\font\titlerm=cmr10 \tfontsize \font\titlerms=cmr7 \tfontsize
\font\titlermss=cmr5 \tfontsize \font\titlei=cmmi10 \tfontsize
\font\titleis=cmmi7 \tfontsize \font\titleiss=cmmi5 \tfontsize
\font\titlesy=cmsy10 \tfontsize \font\titlesys=cmsy7 \tfontsize
\font\titlesyss=cmsy5 \tfontsize \font\titleit=cmti10 \tfontsize
\skewchar\titlei='177 \skewchar\titleis='177 \skewchar\titleiss='177
\skewchar\titlesy='60 \skewchar\titlesys='60 \skewchar\titlesyss='60
\def\titlefont{\def\rm{\fam0\titlerm}
\textfont0=\titlerm \scriptfont0=\titlerms \scriptscriptfont0=\titlermss
\textfont1=\titlei \scriptfont1=\titleis \scriptscriptfont1=\titleiss
\textfont2=\titlesy \scriptfont2=\titlesys \scriptscriptfont2=\titlesyss
\textfont\itfam=\titleit \def\it{\fam\itfam\titleit}\rm}
 \ifx\answ\bigans\else scaled\magstep1\fi
\ifx\answ\bigans\def\abstractfont{\tenpoint}\else
\font\abssl=cmsl10 scaled \magstep1
\font\absrm=cmr10 scaled\magstep1 \font\absrms=cmr7 scaled\magstep1
\font\absrmss=cmr5 scaled\magstep1 \font\absi=cmmi10 scaled\magstep1
\font\absis=cmmi7 scaled\magstep1 \font\absiss=cmmi5 scaled\magstep1
\font\abssy=cmsy10 scaled\magstep1 \font\abssys=cmsy7 scaled\magstep1
\font\abssyss=cmsy5 scaled\magstep1 \font\absbf=cmbx10 scaled\magstep1
\skewchar\absi='177 \skewchar\absis='177 \skewchar\absiss='177
\skewchar\abssy='60 \skewchar\abssys='60 \skewchar\abssyss='60
\def\abstractfont{\def\rm{\fam0\absrm}
\textfont0=\absrm \scriptfont0=\absrms \scriptscriptfont0=\absrmss
\textfont1=\absi \scriptfont1=\absis \scriptscriptfont1=\absiss
\textfont2=\abssy \scriptfont2=\abssys \scriptscriptfont2=\abssyss
\textfont\itfam=\bigit \def\it{\fam\itfam\bigit}\def\footnotefont{\tenpoint}%
\textfont\slfam=\abssl \def\sl{\fam\slfam\abssl}%
\textfont\bffam=\absbf \def\bf{\fam\bffam\absbf}\rm}\fi
\def\tenpoint{\def\rm{\fam0\tenrm}
\textfont0=\tenrm \scriptfont0=\sevenrm \scriptscriptfont0=\fiverm
\textfont1=\teni  \scriptfont1=\seveni  \scriptscriptfont1=\fivei
\textfont2=\tensy \scriptfont2=\sevensy \scriptscriptfont2=\fivesy
\textfont\itfam=\tenit
\def\it{\fam\itfam\tenit}\def\footnotefont{\ninepoint}%
\textfont\bffam=\tenbf \def\bf{\fam\bffam\tenbf}\def\sl{\fam\slfam\tensl}\rm}
\font\ninerm=cmr9 \font\sixrm=cmr6 \font\ninei=cmmi9 \font\sixi=cmmi6
\font\ninesy=cmsy9 \font\sixsy=cmsy6 \font\ninebf=cmbx9
\font\nineit=cmti9 \font\ninesl=cmsl9 \skewchar\ninei='177
\skewchar\sixi='177 \skewchar\ninesy='60 \skewchar\sixsy='60
\def\ninepoint{\def\rm{\fam0\ninerm}
\textfont0=\ninerm \scriptfont0=\sixrm \scriptscriptfont0=\fiverm
\textfont1=\ninei \scriptfont1=\sixi \scriptscriptfont1=\fivei
\textfont2=\ninesy \scriptfont2=\sixsy \scriptscriptfont2=\fivesy
\textfont\itfam=\ninei \def\it{\fam\itfam\nineit}\def\sl{\fam\slfam\ninesl}%
\textfont\bffam=\ninebf \def\bf{\fam\bffam\ninebf}\rm}
%
%

\hyphenation{anom-aly anom-alies coun-ter-term coun-ter-terms}
\def\inv{^{\raise.15ex\hbox{${\scriptscriptstyle -}$}\kern-.05em 1}}

\def\Dsl{\,\raise.15ex\hbox{/}\mkern-13.5mu D} 
\def\dsl{\raise.15ex\hbox{/}\kern-.57em\partial}

\def\tr{{\rm tr}} 
\font\bigit=cmti10 scaled \magstep1
\def\lspace{\ifx\answ\bigans{}\else\qquad\fi}
\def\lbspace{\ifx\answ\bigans{}\else\hskip-.2in\fi} 
\def\boxeqn#1{\vcenter{\vbox{\hrule\hbox{\vrule\kern3pt\vbox{\kern3pt
           \hbox{${\displaystyle #1}$}\kern3pt}\kern3pt\vrule}\hrule}}}
\def\mbox#1#2{\vcenter{\hrule \hbox{\vrule height#2in
               \kern#1in \vrule} \hrule}}  
%

\def\e#1{{\rm e}^{^{\textstyle#1}}}

\def\darr#1{\raise1.5ex\hbox{$\leftrightarrow$}\mkern-16.5mu #1}

\def\roughly#1{\raise.3ex\hbox{$#1$\kern-.75em\lower1ex\hbox{$\sim$}}}



\def\IB{\relax\hbox{$\inbar\kern-.3em{\rm B}$}}
\def\IC{\relax\hbox{$\inbar\kern-.3em{\rm C}$}}
\def\ID{\relax\hbox{$\inbar\kern-.3em{\rm D}$}}
\def\IE{\relax\hbox{$\inbar\kern-.3em{\rm E}$}}
\def\IF{\relax\hbox{$\inbar\kern-.3em{\rm F}$}}
\def\IG{\relax\hbox{$\inbar\kern-.3em{\rm G}$}}
\def\IGa{\relax\hbox{${\rm I}\kern-.18em\Gamma$}}
\def\IH{\relax{\rm I\kern-.18em H}}
\def\IK{\relax{\rm I\kern-.18em K}}
\def\II{\relax{\rm I\kern-.18em I}}
\def\IL{\relax{\rm I\kern-.18em L}}
\def\IP{\relax{\rm I\kern-.18em P}}
\def\IR{\relax{\rm I\kern-.18em R}}
\def\IZ{\relax\ifmmode\mathchoice {\hbox{\cmss Z\kern-.4em Z}}{\hbox{\cmss
Z\kern-.4em Z}} {\lower.9pt\hbox{\cmsss Z\kern-.4em Z}}
{\lower1.2pt\hbox{\cmsss Z\kern-.4em Z}}\else{\cmss Z\kern-.4em Z}\fi}

\def\IB{\relax{\rm I\kern-.18em B}}
\def\IC{{\relax\hbox{$\inbar\kern-.3em{\rm C}$}}}
\def\ID{\relax{\rm I\kern-.18em D}}
\def\IE{\relax{\rm I\kern-.18em E}}
\def\IF{\relax{\rm I\kern-.18em F}}


\def\CW {{\cal W}}


\def\pa{\partial}




\def\s{\lies}


\def\demi{{1\over 2}}


\def\P{\Psi}    
\def\F{\Phi}

\def\a{\alpha}
\def\b{\beta}

\def\m{\mu}
\def\n{\nu}
\def\r{\rho}
 
\def\k{\kappa}
\def\e{\epsilon}

\def\|{\Big|}
\def\({\Big(}   \def\){\Big)}
\def\[{\Big[}   \def\]{\Big]}



\def\paper#1#2#3#4{#1, {\sl #2}, #3 {\tt #4}}

\def\hh{hep-th/}


\def\PLB#1#2#3{Phys. Lett.~{\bf B#1} (#2) #3}
\def\NPB#1#2#3{Nucl. Phys.~{\bf B#1} (#2) #3}
\def\PRL#1#2#3{Phys. Rev. Lett.~{\bf #1} (#2) #3}
\def\CMP#1#2#3{Comm. Math. Phys.~{\bf #1} (#2) #3}
\def\PRD#1#2#3{Phys. Rev.~{\bf D#1} (#2) #3}
\def\MPL#1#2#3{Mod. Phys. Lett.~{\bf #1} (#2) #3}
\def\IJMP#1#2#3{Int. Jour. Mod. Phys.~{\bf #1} (#2) #3}


\def\unlockat{\catcode`\@=11}
\def\lockat{\catcode`\@=12}

\unlockat


\def\newsec#1{\global\advance\secno by1\message{(\the\secno. #1)}
\global\subsecno=0\global\subsubsecno=0\eqnres@t\noindent {\bf\the\secno. #1}
\writetoca{{\secsym} {#1}}\par\nobreak\medskip\nobreak}
\global\newcount\subsecno \global\subsecno=0
\def\subsec#1{\global\advance\subsecno by1\message{(\secsym\the\subsecno.
#1)}
\ifnum\lastpenalty>9000\else\bigbreak\fi\global\subsubsecno=0
\noindent{\it\secsym\the\subsecno. #1}
\writetoca{\string\quad {\secsym\the\subsecno.} {#1}}
\par\nobreak\medskip\nobreak}
\global\newcount\subsubsecno \global\subsubsecno=0
\def\subsubsec#1{\global\advance\subsubsecno by1
\message{(\secsym\the\subsecno.\the\subsubsecno. #1)}
\ifnum\lastpenalty>9000\else\bigbreak\fi
\noindent\quad{\secsym\the\subsecno.\the\subsubsecno.}{#1}
\writetoca{\string\qquad{\secsym\the\subsecno.\the\subsubsecno.}{#1}}
\par\nobreak\medskip\nobreak}

\def\subsubseclab#1{\DefWarn#1\xdef #1{\noexpand\hyperref{}{subsubsection}%
{\secsym\the\subsecno.\the\subsubsecno}%
{\secsym\the\subsecno.\the\subsubsecno}}%
\writedef{#1\leftbracket#1}\wrlabeL{#1=#1}}
\lockat

\def\dbend{\lower3.5pt\hbox{\manual\char127}}


\def\boxit#1{\vbox{\hrule\hbox{\vrule\kern8pt
\vbox{\hbox{\kern8pt}\hbox{\vbox{#1}}\hbox{\kern8pt}}
\kern8pt\vrule}\hrule}}

\def\mathboxit#1{\vbox{\hrule\hbox{\vrule\kern8pt\vbox{\kern8pt
\hbox{$\displaystyle #1$}\kern8pt}\kern8pt\vrule}\hrule}}


\def\inbar{\,\vrule height1.5ex width.4pt depth0pt}

\font\cmss=cmss10 \font\cmsss=cmss10 at 7pt


\lref\simons{ J. Cheeger and J. Simons, {\it Differential Characters and
Geometric Invariants},  Stony Brook Preprint, (1973), unpublished.}

\lref\cargese{ L.~Baulieu, {\it Algebraic quantization of gauge theories},
Perspectives in fields and particles, Plenum Press, eds. Basdevant-Levy,
Cargese Lectures 1983}

\lref\antifields{ L. Baulieu, M. Bellon, S. Ouvry, C.Wallet, Phys.Letters
B252 (1990) 387; M.  Bocchichio, Phys. Lett. B187 (1987) 322;  Phys. Lett. B
192 (1987) 31; R.  Thorn    Nucl. Phys.   B257 (1987) 61. }

\lref\thompson{ George Thompson,  Annals Phys. 205 (1991) 130; J.M.F.
Labastida, M. Pernici, Phys. Lett. 212B  (1988) 56; D. Birmingham, M.Blau,
M. Rakowski and G.Thompson, Phys. Rept. 209 (1991) 129.}

\lref\tonin{ Tonin}

\lref\wittensix{ E.  Witten, {\it New  Gauge  Theories In Six Dimensions},
\hh{9710065}. }

\lref\orlando{ O. Alvarez, L. A. Ferreira and J. Sanchez Guillen, {\it  A New
Approach to Integrable Theories in any Dimension}, hep-th/9710147.}

\lref\wittentopo{ E.  Witten,  {\it  Topological Quantum Field Theory},
\hh9403195, Commun.  Math. Phys.  {117} (1988)353.  }

\lref\wittentwist{ E.  Witten, {\it Supersymmetric Yang--Mills theory on a
four-manifold}, J.  Math.  Phys.  {35} (1994) 5101.}

\lref\west{ L.~Baulieu, P.~West, {\it Six Dimensional TQFTs and  Self-dual
Two-Forms,} Phys.Lett. B {\bf 436 } (1998) 97, /hep-th/9805200}

\lref\bv{ I.A. Batalin and V.A. Vilkowisky,    Phys. Rev.   D28  (1983)
2567\semi M. Henneaux,  Phys. Rep.  126   (1985) 1\semi M. Henneaux and C.
Teitelboim, {\it Quantization of Gauge Systems}
  Princeton University Press,  Princeton (1992).}

\lref\kyoto{ L. Baulieu, E. Bergschoeff and E. Sezgin, Nucl. Phys.
B307(1988)348\semi L. Baulieu,   {\it Field Antifield Duality, p-Form Gauge
Fields
   and Topological Quantum Field Theories}, hep-th/9512026,
   Nucl. Phys. B478 (1996) 431.  }

\lref\sourlas{ G. Parisi and N. Sourlas, {\it Random Magnetic Fields,
Supersymmetry and Negative Dimensions}, Phys. Rev. Lett.  43 (1979) 744;
Nucl.  Phys.  B206 (1982) 321.  }

\lref\SalamSezgin{ A.  Salam  and  E.  Sezgin, {\it Supergravities in
diverse dimensions}, vol.  1, p. 119\semi P.  Howe, G.  Sierra and P.
Townsend, Nucl Phys B221 (1983) 331.}

\lref\nekrasov{ A. Losev, G. Moore, N. Nekrasov, S. Shatashvili, {\it
Four-Dimensional Avatars of Two-Dimensional RCFT},  hep-th/9509151, Nucl.
Phys.  Proc.  Suppl.   46 (1996) 130\semi L.  Baulieu, A.  Losev,
N.~Nekrasov  {\it Chern-Simons and Twisted Supersymmetry in Higher
Dimensions},  hep-th/9707174, to appear in Nucl.  Phys.  B.  }

\lref\WitDonagi{R.~ Donagi, E.~ Witten, ``Supersymmetric Yang--Mills Theory
and Integrable Systems'', hep-th/9510101, Nucl. Phys.{\bf B}460 (1996)
299-334}
\lref\Witfeb{E.~ Witten, ``Supersymmetric Yang--Mills Theory On A
Four-Manifold,''  hep-th/9403195; J. Math. Phys. {\bf 35} (1994) 5101.}
\lref\Witgrav{E.~ Witten, ``Topological Gravity'', Phys.Lett.206B:601, 1988}
\lref\witaffl{I. ~ Affleck, J.A.~ Harvey and E.~ Witten,
        ``Instantons and (Super)Symmetry Breaking
        in $2+1$ Dimensions'', Nucl. Phys. {\bf B}206 (1982) 413}

\lref\Witgravcs{E.~ Witten, {\it Quantization of Chern--Simons Theory with Complex Gauge group},   Commun. Math.Phys. 137  (1991)  29-66}

\lref\wittabl{E.~ Witten,  ``On $S$-Duality in Abelian Gauge Theory,''
hep-th/9505186; Selecta Mathematica {\bf 1} (1995) 383}
\lref\wittgr{E.~ Witten, ``The Verlinde Algebra And The Cohomology Of The
Grassmannian'',  hep-th/9312104}
\lref\wittenwzw{E. Witten, ``Non abelian bosonization in two dimensions,''
Commun. Math. Phys. {\bf 92} (1984)455 }
\lref\witgrsm{E. Witten, ``Quantum field theory, grassmannians and algebraic
curves,'' Commun.Math.Phys.113:529,1988}
\lref\wittjones{E. Witten, ``Quantum field theory and the Jones
polynomial,'' Commun.  Math. Phys., 121 (1989) 351. }
\lref\witttft{E.~ Witten, ``Topological Quantum Field Theory", Commun. Math.
Phys. {\bf 117} (1988) 353.}
\lref\wittmon{E.~ Witten, ``Monopoles and Four-Manifolds'', hep-th/9411102}
\lref\Witdgt{ E.~ Witten, ``On Quantum gauge theories in two dimensions,''
Commun. Math. Phys. {\bf  141}  (1991) 153}
\lref\witrevis{E.~ Witten,
 ``Two-dimensional gauge theories revisited'', hep-th/9204083; J. Geom.
Phys. 9 (1992) 303-368}
\lref\Witgenus{E.~ Witten, ``Elliptic Genera and Quantum Field Theory'',
Comm. Math. Phys. 109(1987) 525. }
\lref\OldZT{E. Witten, ``New Issues in Manifolds of SU(3) Holonomy,'' {\it
Nucl. Phys.} {\bf B268} (1986) 79 \semi J. Distler and B. Greene, ``Aspects
of (2,0) String Compactifications,'' {\it Nucl. Phys.} {\bf B304} (1988) 1
\semi B. Greene, ``Superconformal Compactifications in Weighted Projective
Space,'' {\it Comm. Math. Phys.} {\bf 130} (1990) 335.}
\lref\bagger{E.~ Witten and J. Bagger, Phys. Lett. {\bf 115B}(1982) 202}
\lref\witcurrent{E.~ Witten,``Global Aspects of Current Algebra'',
Nucl.Phys.B223 (1983) 422\semi ``Current Algebra, Baryons and Quark
Confinement'', Nucl.Phys. B223 (1993) 433}
\lref\Wittreiman{S.B. Treiman, E. Witten, R. Jackiw, B. Zumino, ``Current
Algebra and Anomalies'', Singapore, Singapore: World Scientific ( 1985) }
\lref\Witgravanom{L. Alvarez-Gaume, E.~ Witten, ``Gravitational Anomalies'',
Nucl.Phys.B234:269,1984. }

\lref\nicolai{\paper {H.~Nicolai}{New Linear Systems for 2D Poincar\'e
Supergravities}{\NPB{414}{1994}{299},}{\hh 9309052}.}



\lref\baex{\paper {L.~Baulieu, B.~Grossman}{Monopoles and Topological Field
Theory}{\PLB{214}{1988}{223}.}{}\paper {L.~Baulieu}{Chern-Simons Three-Dimensional and
Yang--Mills-Higgs Two-Dimensional Systems as Four-Dimensional Topological
Quantum Field Theories}{\PLB{232}{1989}{473}.}{}}

\lref\bg{\paper {L.~Baulieu, B.~Grossman}{Monopoles and Topological Field
Theory}{\PLB{214}{1988}{223}.}{}}

\lref\seibergsix{\paper {N.~Seiberg}{Non-trivial Fixed Points of The
Renormalization Group in Six
 Dimensions}{\PLB{390}{1997}{169}}{\hh 9609161}\semi
\paper {O.J.~Ganor, D.R.~Morrison, N.~Seiberg}{
  Branes, Calabi-Yau Spaces, and Toroidal Compactification of the N=1
  Six-Dimensional $E_8$ Theory}{\NPB{487}{1997}{93}}{\hh 9610251}\semi
\paper {O.~Aharony, M.~Berkooz, N.~Seiberg}{Light-Cone
  Description of (2,0) Superconformal Theories in Six
  Dimensions}{Adv. Theor. Math. Phys. {\bf 2} (1998) 119}{\hh 9712117.}}

\lref\cs{\paper {L.~Baulieu}{Chern-Simons Three-Dimensional and
Yang--Mills-Higgs Two-Dimensional Systems as Four-Dimensional Topological
Quantum Field Theories}{\PLB{232}{1989}{473}.}{}}

\lref\beltrami{\paper {L.~Baulieu, M.~Bellon}{Beltrami Parametrization and
String Theory}{\PLB{196}{1987}{142}}{}\semi
\paper {L.~Baulieu, M.~Bellon, R.~Grimm}{Beltrami Parametrization For
Superstrings}{\PLB{198}{1987}{343}}{}\semi
\paper {R.~Grimm}{Left-Right Decomposition of Two-Dimensional Superspace
Geometry and Its BRS Structure}{Annals Phys. {\bf 200} (1990) 49.}{}}

\lref\bbg{\paper {L.~Baulieu, M.~Bellon, R.~Grimm}{Left-Right Asymmetric
Conformal Anomalies}{\PLB{228}{1989}{325}.}{}}

\lref\bonora{\paper {G.~Bonelli, L.~Bonora, F.~Nesti}{String Interactions
from Matrix String Theory}{\NPB{538}{1999}{100},}{\hh 9807232}\semi
\paper {G.~Bonelli, L.~Bonora, F.~Nesti, A.~Tomasiello}{Matrix String Theory
and its Moduli Space}{}{\hh 9901093.}}

\lref\corrigan{\paper {E.~Corrigan, C.~Devchand, D.B.~Fairlie,
J.~Nuyts}{First Order Equations for Gauge Fields in Spaces of Dimension
Greater Than Four}{\NPB{214}{452}{1983}.}{}}

\lref\acha{\paper {B.S.~Acharya, M.~O'Loughlin, B.~Spence}{Higher
Dimensional Analogues of Donaldson-Witten Theory}{\NPB{503}{1997}{657},}{\hh
9705138}\semi
\paper {B.S.~Acharya, J.M.~Figueroa-O'Farrill, M.~O'Loughlin,
B.~Spence}{Euclidean
  D-branes and Higher-Dimensional Gauge   Theory}{\NPB{514}{1998}{583},}{\hh
  9707118.}}

\lref\Witr{\paper{E.~Witten}{Introduction to Cohomological Field   Theories}
{Lectures at Workshop on Topological Methods in Physics (Trieste, Italy, Jun
11-25, 1990), \IJMP{A6}{1991}{2775}.}{}}

\lref\ohta{\paper {L.~Baulieu, N.~Ohta}{Worldsheets with Extended
Supersymmetry} {\PLB{391}{1997}{295},}{\hh 9609207}.}

\lref\gravity{\paper {L.~Baulieu}{Transmutation of Pure 2-D Supergravity
Into Topological 2-D Gravity and Other Conformal Theories}
{\PLB{288}{1992}{59},}{\hh 9206019.}}

\lref\wgravity{\paper {L.~Baulieu, M.~Bellon, R.~Grimm}{Some Remarks on  the
Gauging of the Virasoro and   $w_{1+\infty}$
Algebras}{\PLB{260}{1991}{63}.}{}}

\lref\fourd{\paper {E.~Witten}{Topological Quantum Field
Theory}{\CMP{117}{1988}{353}}{}\semi
\paper {L.~Baulieu, I.M.~Singer}{Topological Yang--Mills Symmetry}{Nucl.
Phys. Proc. Suppl. {\bf 15B} (1988) 12.}{}}

\lref\topo{\paper {L.~Baulieu}{On the Symmetries of Topological Quantum Field
Theories}{\IJMP{A10}{1995}{4483},}{\hh 9504015}\semi
\paper {R.~Dijkgraaf, G.~Moore}{Balanced Topological Field
Theories}{\CMP{185}{1997}{411},}{\hh 9608169.}}

\lref\wwgravity{\paper {I.~Bakas} {The Large $N$ Limit   of Extended
Conformal Symmetries}{\PLB{228}{1989}{57}.}{}}

\lref\wwwgravity{\paper {C.M.~Hull}{Lectures on $\CW$-Gravity,
$\CW$-Geometry and
$\CW$-Strings}{}{\hh 9302110}, and~references therein.}

\lref\wvgravity{\paper {A.~Bilal, V.~Fock, I.~Kogan}{On the origin of
$W$-algebras}{\NPB{359}{1991}{635}.}{}}

\lref\surprises{\paper {E.~Witten} {Surprises with Topological Field
Theories} {Lectures given at ``Strings 90'', Texas A\&M, 1990,}{Preprint
IASSNS-HEP-90/37.}}

\lref\stringsone{\paper {L.~Baulieu, M.B.~Green, E.~Rabinovici}{A Unifying
Topological Action for Heterotic and  Type II Superstring  Theories}
{\PLB{386}{1996}{91},}{\hh 9606080.}}

\lref\stringsN{\paper {L.~Baulieu, M.B.~Green, E.~Rabinovici}{Superstrings
from   Theories with $N>1$ World Sheet Supersymmetry}
{\NPB{498}{1997}{119},}{\hh 9611136.}}

\lref\bks{\paper {L.~Baulieu, H.~Kanno, I.~Singer}{Special Quantum Field
Theories in Eight and Other Dimensions}{\CMP{194}{1998}{149},}{\hh
9704167}\semi
\paper {L.~Baulieu, H.~Kanno, I.~Singer}{Cohomological Yang--Mills Theory
  in Eight Dimensions}{ Talk given at APCTP Winter School on Dualities in
String Theory (Sokcho, Korea, February 24-28, 1997),} {\hh 9705127.}}

\lref\witdyn{\paper {P.~Townsend}{The eleven dimensional supermembrane
revisited}{\PLB{350}{1995}{184},}{\hh9501068}\semi
\paper{E.~Witten}{String Theory Dynamics in Various Dimensions}
{\NPB{443}{1995}{85},}{\hh 9503124}.}

\lref\bfss{\paper {T.~Banks, W.Fischler, S.H.~Shenker,
L.~Susskind}{$M$-Theory as a Matrix Model~:
A~Conjecture}{\PRD{55}{1997}{5112},} {\hh9610043.}}

\lref\seiberg{\paper {N.~Seiberg}{Why is the Matrix Model
Correct?}{\PRL{79}{1997}{3577},} {\hh 9710009.}}

\lref\sen{\paper {A.~Sen}{$D0$ Branes on $T^n$ and Matrix Theory}{Adv.
Theor. Math. Phys.~{\bf 2} (1998) 51,} {\hh 9709220.}}

\lref\laroche{\paper {L.~Baulieu, C.~Laroche} {On Generalized Self-Duality
Equations Towards Supersymmetric   Quantum Field Theories Of
Forms}{\MPL{A13}{1998}{1115},}{\hh  9801014.}}

\lref\bsv{\paper {M.~Bershadsky, V.~Sadov, C.~Vafa} {$D$-Branes and
Topological Field Theories}{\NPB{463} {1996}{420},}{\hh9511222.}}

\lref\vafapuzz{\paper {C.~Vafa}{Puzzles at Large N}{}{\hh 9804172.}}

\lref\dvv{\paper {R.~Dijkgraaf, E.~Verlinde, H.~Verlinde} {Matrix String
Theory}{\NPB{500}{1997}{43},} {\hh9703030.}}

\lref\wynter{\paper {T.~Wynter}{Gauge Fields and Interactions in Matrix
String Theory}{\PLB{415}{1997}{349},}{\hh9709029.}}

\lref\kvh{\paper {I.~Kostov, P.~Vanhove}{Matrix String Partition
Functions}{}{\hh9809130.}}

\lref\ikkt{\paper {N.~Ishibashi, H.~Kawai, Y.~Kitazawa, A.~Tsuchiya} {A
Large $N$ Reduced Model as Superstring}{\NPB{498} {1997}{467},}{\hh
9612115.}}

\lref\ss{\paper {S.~Sethi, M.~Stern} {$D$-Brane Bound States
Redux}{\CMP{194}{1998} {675},}{\hh 9705046.}}

\lref\mns{\paper {G.~Moore, N.~Nekrasov, S.~Shatashvili} {$D$-particle Bound
States and Generalized Instantons}{} {\hh 9803265.}}

\lref\bsh{\paper {L.~Baulieu, S.~Shatashvili} {Duality from Topological
Symmetry}{} {\hh 9811198.}}

\lref\pawu{ {G.~Parisi, Y.S.~Wu} {}{ Sci. Sinica  {\bf 24} {(1981)} {484}.}}

\lref\coulomb{ {L.~Baulieu, D.~Zwanziger, }   {\it Renormalizable Non-Covariant
Gauges and Coulomb Gauge Limit}, {Nucl.Phys. B {\bf 548 } (1999) 527.} {\hh
9807024}.}

\lref\rcoulomb{ {D.~Zwanziger, }   {\it Renormalization in the Coulomb
gauge and order parameter for confinement in QCD}, {Nucl.Phys. B {\bf 538
} (1998) 237.} {}}

\lref\horne{ {J.H.~Horne, }   {\it
Superspace versions of Topological Theories}, {Nucl.Phys. B {\bf 318
} (1989) 22.} {}}

\lref\sto{ {S.~Ouvry, R.~Stora, P.~Van~Baal }   {\it
}, {Phys. Lett. B {\bf 220
} (1989) 159;} {}{ R.~Stora, {\it Exercises in   Equivariant Cohomology},
In Quabtum Fields and Quantum Space Time, Edited 
by 't Hooft et al., Plenum Press, New York, 1997}            }

\lref\dzvan{ {D.~Zwanziger, }   {\it Vanishing of zero-momentum lattice
gluon propagator and color confinement}, {Nucl.Phys. B {\bf 364 } (1991)
127.} }

\lref\dan{ {D.~Zwanziger},  {\it Covariant Quantization of Gauge
Fields without }{Nucl. Phys. B {\bf   192}, (1981) {259}.}{}}

\lref\danzinn{  {J.~Zinn-Justin, D.~Zwanziger, } {}{Nucl. Phys. B  {\bf
295} (1988) {297}.}{}}

\lref\danlau{ {L.~Baulieu, D.~Zwanziger, } {\it Equivalence of Stochastic
Quantization and the-Popov Ansatz,
  }{Nucl. Phys. B  {\bf 193 } (1981) {163}.}{}}

\lref\munoz{ { A.~Munoz Sudupe, R. F. Alvarez-Estrada, } {}
Phys. Lett. {\bf 164} (1985) 102; {} {\bf 166B} (1986) 186. }

\lref\okano{ { K.~Okano, } {}
Nucl. Phys. {\bf B289} (1987) 109; {} Prog. Theor. Phys.
suppl. {\bf 111} (1993) 203. }

\lref\singer{
 I.M. Singer, { Comm. of Math. Phys. {\bf 60} (1978) 7.}}

\lref\neu{ {H.~Neuberger,} {Phys. Lett. B {\bf 295}
(1987) {337}.}{}}

\lref\testa{ {M.~Testa,} {}{Phys. Lett. B {\bf 429}
(1998) {349}.}{}}

\lref\Martin{ L.~Baulieu and M. Schaden, {\it Gauge Group TQFT and Improved
Perturbative Yang--Mills Theory}, {  Int. Jour. Mod.  Phys. A {\bf  13}
(1998) 985},   hep-th/9601039.}

\lref\baugrav{  L.~Baulieu, A.~Bilal,  M.~Picco, {\it 
Stochastic Quantization of 2-D Gravity and its link with  3-D Gravity
and  Topological 4-D Gravity}, Nucl.Phys. B {\bf 346}  (1990) 507.}

\lref\baugros{ {L.~Baulieu, B.~Grossman, } {\it A topological Interpretation
of  Stochastic Quantization} {Phys. Lett. B {\bf  212} {(1988)} {351}.}}

\lref\bautop{ {L.~Baulieu}{ \it Stochastic and Topological Field Theories},
{Phys. Lett. B {\bf   232} (1989) {479}}{}; {}{ \it Topological Field Theories
And Gauge Invariance in Stochastic Quantization}, {Int. Jour. Mod.  Phys. A
{\bf  6} (1991) {2793}.}{}}

\lref\bautopr{  {L.~Baulieu, B.~Grossman, } {\it A topological Interpretation
of  Stochastic Quantization} {Phys. Lett. B {\bf  212} {(1988)} {351}};
 {L.~Baulieu}{ \it Stochastic and Topological Field Theories},
{Phys. Lett. B {\bf   232} (1989) {479}}{}; {}{ \it Topological Field Theories
And Gauge Invariance in Stochastic Quantization}, {Int. 
Jour. Mod.  Phys. 
{\bf A6} (1991) {2793}.}{}}

\lref\samson{ {L.~Baulieu, S.L.~Shatashvili, { \it Duality from Topological
Symmetry}, {JHEP {\bf 9903} (1999) 011, hep-th/9811198.}}}{}

\lref\halpern{ {H.S.~Chan, M.B.~Halpern}{}, {Phys. Rev. D {\bf   33} (1986)
{540}.}}

\lref\yue{ {Yue-Yu}, {Phys. Rev. D {\bf   33} (1989) {540}.}}

\lref\neuberger{ {H.~Neuberger,} {\it Non-perturbative gauge Invariance},
{ Phys. Lett. B {\bf 175} (1986) {69}.}{}}

\lref\gribov{  {V.N.~Gribov,} {}{Nucl. Phys. B {\bf 139} (1978) {1}.}{}}

\lref\huffel{ {P.H.~Daamgard, H. Huffel},  {}{Phys. Rep. {\bf 152} (1987)
{227}.}{}}

\lref\creutz{ {M.~Creutz},  {\it Quarks, Gluons and  Lattices, }  Cambridge
University Press 1983, pp 101-107.}

\lref\zinn{ {J. ~Zinn-Justin, }  {Nucl. Phys. B {\bf  275} (1986) {135}.}}

\lref\shamir{  {Y.~Shamir,  } {\it Lattice Chiral Fermions
  }{ Nucl.  Phys.  Proc.  Suppl.  {\bf } 47 (1996) 212,  hep-lat/9509023;
V.~Furman, Y.~Shamir, Nucl.Phys. B {\bf 439 } (1995), hep-lat/9405004.}}

 \lref\kaplan{ {D.B.~Kaplan, }  {\it A Method for Simulating Chiral
Fermions on the Lattice,} Phys. Lett. B {\bf 288} (1992) 342; {\it Chiral
Fermions on the Lattice,}  {  Nucl. Phys. B, Proc. Suppl.  {\bf 30} (1993)
597.}}

\lref\neubergerr{ {H.~Neuberger, } {\it Chirality on the Lattice},
hep-lat/9808036.}

\lref\zbgr {L.~Baulieu, D. Zwanziger, {\it QCD$_4$ From a
Five-Dimensional Point of View},    hep-th/9909006, to appear in Nucl.
Phys. B; ibid,   in peparation; L.~Baulieu, P.A~Grassi,  D. Zwanziger,{
\it Gauge and Topological Symmetries in the Bulk Quantization of
Gauge Theories },     hep-th/9909006.}

\lref\friedan {D.~Friedan , {\it Non Linear Models in $2+\epsilon$
Dimensions},  Annals Phys. {\bf 163}   (1985)  318.    }
  
\lref\hverlinde{ E.~Verlinde, {\it On RG-flow and the Cosmological Constant},
Class. Quant. Grav. 17 (2000) 1277, hep-th/9912058; 
E.~Verlinde, H.~Verlinde,  {\it Gravity and the Cosmological Constant},
hep-th/9912018}

\lref\tian { X.H~ Zhu, G.~Tian, {\it Uniqueness of
Kahler-Ricci solitons on compact
Kdhler manifolds}, C. R. Acad. Sci. Paris Sir. 
I Math. 329 (1999), no.
11, 991--995.}


\Title{\vbox
{\baselineskip 10pt
\hbox{hep-th/0007027}
\hbox{LPTHE-00-27}
 \hbox{   }
}}
{\vbox{\vskip -30 true pt
\centerline{
   }
\medskip
 \centerline{A Curious Relation Between Gravity and Yang--Mills Theories
  }
\medskip
\vskip4pt }}
\centerline{{\bf Laurent Baulieu}$^{\star \dag  \S    }$
   }
\centerline{baulieu@lpthe.jussieu.fr  }
\vskip 0.5cm
\centerline{\it $^{\star}$LPTHE, Universit{\'e}s P. \& M. Curie (Paris~VI) et
D. Diderot (Paris~VII), Paris,  France,}

\centerline{\it $^{\dag}$ Enrico Fermi Inst. and Dept. of Physics, University
of Chicago, Chicago, IL 60637, USA }
\vskip  1cm
 \centerline{\bf  Talk given in the memory of E.S. Fradkin }

\medskip 
\vskip 1cm 
\noindent 
We find that Euclidian or Minkowski gravity in $d$ dimensions can be
formally expressed as the restriction to a slice of a supersymmetric
Yang--Mills theory in $d+1$ dimensions with $SO(d+1)$, $SO(d,1)$ or
$SO(d-1,2)$ internal symmetry. We suggest that renormalization effects in
the bulk imply a contraction of the latter symmetry into the Poincar\'e
group $ISO(d)$ or $ISO(d-1,1)$.

\Date{\ }

\def\e{\epsilon}
\def\demi{{1\over 2}}

\def\pa{\partial}
\def\a{\alpha}
\def\b{\beta}

\def\m{\mu}
\def\n{\nu}
\def\r{\rho}
\def\s{\sigma}

\def\k{\kappa}
\def\P{\Psi}
\def\F{\Phi}

\newsec{Introduction}

   There are conceptual  advantages  for  defining   the      
renormalizable  gauge theories  that describe
four-dimensional physics  from a   a  five-dimensional point of
view. For instance, this  gives    a local quantum field
theory  that is  is  perturbatively renormalizable  by power counting and  
free of the Gribov ambiguity. It also  elucidates several questions about
chirality       \zbgr.  The fifth dimension plays the role of a   
regulator and it can be interpreted as a ``bulk" coordinate that
generalizes the idea of stochastic time.

These progress suggest   investigating the possibility of  embedding 
Einstein $d$-gravity in a  $d+1$ dimensional theory.  Many attempts 
 for  relating   gravity and gauge  theories for the Poincar\'e
group  have been done in the past. They were
never too successful, because of the failure of getting unitary
theories.   Here we   show an  alternative 
 step in
this direction. By combining  the ideas of \zbgr\ and \bautopr, and
using  the property  that Einstein equations are   first order  
in the Cartan formulation, we will 
relate the gravity action in $d$ dimensions to  
a supersymmetric   Yang--Mills theory in $d+1$ dimensions.
The internal gauge symmetry of this bulk theory is either $ISO(d)$ or  $ISO(d-1,1)$, depending if one considers  the Euclidian or  Minkowski case.
 We   suggest that each one of the later symmetries
 can be obtained by a contraction of
 $SO(d+1)$, $SO(d,1)$ or $SO(d-1,2)$  gauge
symmetry  by singular
renormalizations in the ($d+1)$-dimensional bulk theory.   
The
dynamical effects that could be  responsible for such contractions are yet to be
discovered. Unitarity is not a relevant property   for the bulk theory that we consider.  Rather, in the context of bulk quantization, unitarity should emerges  as a
property of   Green functions that are concentrated     in any  given slice of the
$(d+1)$-dimensional space.   
  
The link \bautopr\ between stochastic quantization, whose  Euclidian additional 
time  is a possible interpretation of the internal bulk    coordinate, and 
topological quantum field theory (TQFT) 
  encourages us  to     first consider       
a  TQFT for gravity, using             the metrics   $g_{\mu \nu}$ as a fundamental field.  This actually implies    introducing      an additional  vector field $A^\mu$ to obtain a correct
counting of the degrees of freedom.   The ``self-duality"  equation that
defines  the TQFT  is   the improved Einstein equation introduced
by Friedan \friedan. Solutions of this equation, called Einstein--Ricci
solitons, have been studied in
\tian. The link to renormalization group of $\sigma$-models is quite
appealing, and the bulk internal coordinate looks as the renormalization
scale or the cut-off.

The additional   field $A^\m$ has two possible interpretations. 
It can be viewed as the  expression of  trivial reparametrizations for 
the couplings of  an  underlying $\sigma$ models, as    in
\friedan; on the other hand, in the context of bulk quantization, it 
is merely understood as $g^{\mu,d+1}$, that is,  the component of the
metric
 along the bulk internal radius. It    allows 
for   a theory that is covariant in $d+1$ dimensions. Moreover, one can introduce another independent  field that plays the role of the metrics component $g^{d+1,d+1}$. It expresses the possibility of changing the rate of convergence of correlations functions toward their values in a given slice, without affecting the value of this limit. In the language of stochastic quantization, this means the introduction of a kernel that multiplies the sum of the  drift force and of the noise. Topological invariance in bulk quantization means invariance of observables with respect to   variations of  $g^{\m,d+1}$ and $g^{d+1,d+1}$.

 To see differently   that    gravity in $d$-dimensions is
related to a theory that is covariant in $d+1$ dimensions,  we 
find it natural to  turn
to first order formalism (vielbein formalism).
We will 
exhibit  the  intriguing  curious   feature that we mentioned above, namely a relation between  gravity theory
in
$d$ dimension and a  supersymmetric Yang--Mills theory defined in $d+1$
dimensions.  Indeed,   using   the
Chern--Simons like appearance of Einstein action in first order
formalism, we will find that the bulk formulation of gravity  gives quite
naturally a   theory of the Yang--Mills type with internal $ISO(d) $
or 
$ISO(d-1,1) $ gauge group.

We then remember that 
$ISO(d)$ is  a contraction of    either  
$SO(d,1)$ or $SO(d+1)$. This is a      simplest case.
Indeed, 
 one can choose  $d$
generators  of  
$SO(d+1)$ or $SO(d,1)$, say,  
$M^{d+1,a}$ with   $ 1\leq a \leq d$,       
 redefine  
$ M^{d+1,a}\to P^a=  \lambda\times M^{d+1,a}$  and  take   the limit $\lambda\to \infty$. In this   limit,
 the commutation relations of the $P^a$ and $M^{ a,b}$ are just those
   of  the Poincar\'e group
$ISO(d)$. 
 A    passive point of view   illustrates well  the contraction of  the
$SO(d+1)$ symmetry, acting
 in a space with $d+1$ dimensions, to the
$ISO(d)$ symmetry, acting  in a space with $d$ dimensions.
An observer who is attached  to  the latter space   and can only detects effects of order
$1/\lambda$ will   consider   the $ISO(d)$ symmetry as the natural
symmetry of his space. For similar reasons, $ISO(d-1,1)$ is  a contraction of    either  
$SO(d,1)$ or $SO(d-1,2)$.

  We   find it quite suggestive  that the internal $	ISO(d+1)$ or $ISO(d,1)$
    gauge symmetry of the bulk theory   can be obtained by contractions of 
$SO(d+1)$ or  $SO(d,1)$ for the case of   Euclidian  gravity, or 
$SO(d-1,1)$ or  $SO(d-1,2)$ for the  Minkowski case.  Having a gauge theory in a $(d+1)$-dimensional space with such a gauge symmetry is appealing for describing $d$-gravity. Since   
contractions in a Lie algebra consist  of     
  singular redefinitions of its  generators   that eventually 
satisfy the commutation relations of another    Lie algebra, we
  invoke   dynamical effects in the bulk   that 
 would  renormalize   the structure coefficients  of the bulk internal symmetry   into those of 
the Poincar\'e symmetry in a slice with $d$ dimensions. Taking the limit $\lambda \to \infty$ could be a natural feature in the context of bulk quantization.

  The   range of variation of the internal bulk coordinate   is 
non compact. Starting at
$x^{d+1}=-\infty$, we can stop in any given slice, (not necessarily in  the
boundary at $x^{d+1}= \infty$).   The topological supersymmetry of the theory should ensure that   all possible redefinitions of the ``invisible" bulk
time do not affect observables. In this construction, the theory in $d+1$ dimensions may appear as a sum over equivalent representations of the theory in   
 $d $ dimensions.

 This construction   of  
  a ($d+1$)-dimensional TQFT that determines a 
$d$-dimensional unitary QFT in any given slice,  could realize the   
idea that   the observed Poincar\'e  invariance of theories in $d$-space     results
from their  embedding in a larger  space.    It is only when the
TQFT determines    the physical theory in  a $d$-dimensional  slice    that unitarity is   reached.   The  TQFT  in the
bulk has presumably  a better ultraviolet behavior than gravity, but one
expects subtleties in the limit of concentrating the observables in a
slice. Since the physical $d$-space now appears as  a slice, and not
 as  a boundary, of the 
$(d+1)
$-space,   other transgressions between $d$ and $d+p$ dimensions
seem possible, with interesting perspectives for understanding descent
equations and  anomalies for the special value $p=2$. The idea that a contraction of the   
 $SO(d+1)$,  $SO(d,1)$, or  $SO(d-1, 2)$ invariances occurs because of non perturbative effects  in the bulk  is much more speculative.

 \newsec{TQFT for gravity and renormalization group equation}
\def\tr{{\rm tr}}

 \subsec{The TQFT equation}

Let us first see how one can    construct a TQFT   for gravity in
$d$ dimensions.  
 A possibility is that such a TQFT is related by twist   to
supergravity with N=2 supersymmetry. However,   N=2 supergravity is 
a technically  complicated theory, and we are unfortunately far from understanding the   twist operation for a gravitino. We 
 find it more  realistic  to remain in the pure gravity framework, and 
to  
determine what are the possible  gravitational  gauge functions that   can determine
a TQFT, generalizing the idea of \laroche.   Such a TQFT is   interesting by itself, independently of supergravity. It is quite a challenge to propose a correspondence between the     ghost  spectrum that we find and   twisted  supergravity multiplets.

The topological gauge functions  should   involve in a linear way  the natural 
curvature  in gravity, that is,    the
Ricci tensor
$R _{\mu
\nu}$. This tensor has   the same number of  components as the
 metric  tensor $g_{\mu \nu}$.  However,  a TQFT with a gauge invariance must involve   as
many   gauge fixing-functions as there gauge independent degrees of
freedom in its basic fields \laroche.
Thereby,  we must introduce a vector field $A^\mu$ in addition to
$g_{\mu\nu}$ to obtain a right counting. Indeed,  the fields 
$g_{\mu\nu}$ and $A^{\mu}$ count  for  $d(d+1)/2+d$ degrees of freedom,
that is,    $d(d+1)/2$ degrees of freedom modulo  reparametrization
invariance. This  fits 
 precisely with the number of independent equations  for a
systems of equations linear in    
$R _{\mu \nu}  $. 

We thus introduce    as reparametrization covariant  ``gauge functions'' 
for topological gravity in $d$ dimensions the  $d(d+1)/2$ independent
combinations:
\eqn\gf{\eqalign{&H _{\mu \nu}= R _{\mu \nu} -\demi g _{\mu \nu} R +\k g
_{\mu \nu}+ \nabla  _{\{\m } A_{\n\}} }} We have $R _{\mu \nu}=g
^{\r\s}R_{\r\mu\s \nu} $ and 
$R =g ^{\r\s}R_{\r \s  } $. $\nabla  _\m$ is the covariant derivative.
$\k$ is like a cosmological constant.

The square   of the gauge function
\gf\ gives:
\eqn\gfc{\eqalign{& H _{\mu \nu}H ^{\mu \nu}=( R ^{\mu \nu}-\demi g
^{\mu \nu} R) (R _{\mu \nu} -\demi g _{\mu \nu} R)  -\k R +\k^2 +
\nabla  _{\{\m } A_{\n\}}
\nabla  ^{\{\m } A^{\n\}} }}

When $H _{\mu \nu}$ vanishes, \gf\ gives  the modified Einstein  equation
that   Friedan introduced   to determine the fixed point of the 
$\s$-models \friedan. The equation $H _{\mu \nu}=0$ implies   that 
$\nabla  ^\m \nabla  _{\{\m } A_{\n\}}$ is zero, because of   the Bianchi identity. It  is quite similar to the Bogomolny equation   $F_i=D_i\phi$ that  implies the scalar
field equation of motion $D^iD_i\phi=0$ because of the Bianchi identity for the electric field $F_i$. 
 Solutions  to \gf\ for $H _{\mu
\nu}=0$   are called Ricci solitons \tian. They   minimize  the action \gfc. 

 In the context of a TQFT,      \gfc\
can be seen as the bosonic part of an action   obtained by a usual    
TQFT construction, using
$d(d+1)/2$ of the $d(d+3)/2$  freedoms contained in the set of fields  $g
_{\m \n }$ and $A^\m $ for imposing \gf; the remaining  $d$ freedoms can
be fixed by   a   gauge functions $H^\mu$ that resembles   t'Hooft-Feynman gauges in the spontaneously broken Yang--Mills theory and   fixes the reparametrization invariance in  \gfc:  
\eqn\gfd{\eqalign{& 
 H^\mu =\pa_{ \n } g ^{\m \n }-\a A^\m }} Here $\a$ is a parameter. The
combination of \gf\ and \gfd\  exhausts   all  gauge freedom
contained in the topological invariances for $g _{\m \n }$ and $A^\m $,
and one expects a   TQFT with an  equivariant cohomology with respect to
  reparametrization invariance.  

Where would be  the fermions of this TQFT? They can be  introduced  
by defining the BRST variations of  $g _{\m \n }$ and $A^\m $, which are
arbitrary transformations, defined modulo reparametrization.  $c^\m$ is
the vector ghost for the latter transformations, and $\P_{\m\n}$ and
$\P^\m$ are the anticommuting  topological ghosts for $g _{\m \n }$ and $A^\m
$.
$\F^\m$ is the commuting vector  ghost of ghost for reparametrizations.
\eqn\gfc{\eqalign{   sg _{ \m \n} &=\P _{\m \n } + g _{ \m \r}\pa_\n
c^\r  +g _{ \n \r}\pa_\m c^\r +g _{ \m \n}\pa   _\r c^\r
\cr s \P _{\m \n } &= g _{ \m \r}\pa_\n \F^\r  +g _{ \n \r}\pa_\m \F^\r
+g _{ \m \n}\pa   _\r \F^\r  + \P_{ \m \r}\pa_\n c^\r  +\P _{ \n
\r}\pa_\m c^\r +\P _{ \m \n}\pa   _\r c^\r\cr sc^\m &=
-\F^\m +c^\n\pa_\n c^\m\cr s\F^\m &= c^\n\pa_\n \F^\m -\F^\n\pa_\n c^\m\cr
s A^\mu & =\Psi^\mu + \{A^\mu,c^\mu \}\cr
s \Psi ^\mu & =   \{A^\mu,\Phi^\mu\}-\{\Psi^\mu,c^\mu \}
}}

Due to   reparametrization invariance , we must introduce a
reparametrization covariant gauge function for the topological ghost of
$g_{\m\n}$. It is:
  \eqn\gfc{\eqalign{& 
 \nabla  _{ \m } \P ^{\m \n } }}

To enforce the gauge functions, we need antighosts and Lagrange
multipliers. They are the symmetric tensor $\bar \P^ {\{\m\n\}}$, the
vector 
$\bar c^ {\m}$, and their  Lagrange multipliers $H^ {\{\m\n\}}$ and $ H^ {\m}$,  with 
$s\bar \P^ {\{\m\n\}} =   H^ {\{\m\n\}}$ and  
$ s \bar c^ {\m}=  H^ {\m}$. We have the vector antighosts for antighost
$\bar \F^ {\m}$ and the  fermionic
 Lagrange multiplier
 $\eta^ {\m}$ for \gfc\ and $ s \bar \F^{\m}=  \eta^ {\m}$.

As for the   fermionic degrees of freedom, the set of anticommuting
ghosts and antighosts for the equivariant cohomology is:
\eqn\se{\eqalign{  
 \P^{\{\m\n\}}, \P^\m, \bar \P^{\{\m\n\}} ,\eta^\m
 }} In the sector of the gauge fixing of reparametrization
invariance, it is:
\eqn\sgf{\eqalign{   
\bar c^\m, c^\m
 }}
 The TQFT Lagrangian is:
\eqn\gimp{\eqalign{s\Big ( &\bar \P^{\{\m\n\}}\big ( 
  R _{\mu \nu} -\demi g _{\mu \nu} R+ \nabla  _{\{\m } A_{\n\}} +\k g
_{\mu \nu} +\demi H _{\{\m\n\}}\big) 
\cr &+\bar \F_\n  \nabla  _{ \m } \P ^{\m \n }\cr & \bar c_\n \big ( \pa
_{ \m } g ^{\m \n }-\alpha A^\nu+\demi H ^{\m} \big)
\Big ) }}

  There are actually several points of interest that we decide not
to explore in this paper. We could  introduce   a $w$ symmetry as in \zbgr\ for   having an equivariant BRST symmetry and  disentangle the  topological symmetry and the reparametrization invariance. This also     determines   the action \gimp\ as part of the cohomology of the $w$-symmetry.  Then, 
there is perhaps  a correspondence between  the fermionic ghosts and the
gravitini of some $N=2$ supergravity. Such a correspondence between tensor  ghosts and physical spinors 
is known for the  superstring 
and the super Yang--Mills theories, which  can be twisted into topological theories. 
Logically,  the same situation might occur for supergravity. Finally,  it is likely
that  ``observables'' could be  constructed, by using   descent
equations from the Euler and Pontryagin  invariants. There are
  actually  new  invariants that are   studied by mathematicians
\tian\ and the gravitational  TQFT  could give their interpretation.  We skip those questions, since we are mostly interested by
exploring the perspective offered by \gf\ in the context of bulk
quantization \zbgr.

\subsec{The Renormalization group equation and the stochastic equation}

The TQFT defined by \gf\ contains a term that is quartic in the graviton
propagator.  Its  has an improved ultraviolet behavior  as compared to the situation in ordinary
gravity.   If one tries to directly    get a
particle interpretation, one is readily impeached by the lack of  unitarity due to  spin-$2$
tachyons coming from the quartic propagators.  But,  this contradiction disappears by   considering
  one coordinates as an internal  bulk
coordinate: by definition,  the  TQFT  in the bulk  with dimension $d+1$ does not requite  unitarity
properties; rather  it is only  used  for the bulk
quantization of gravity in one less dimensions
$d$. Defining gravity  as leaving in a
slice, or in the slice of a slice, of an enlarged space, certainly opens new perspectives and unitarity is only required for this restriction of the bulk theory.

 Let indeed introduce another coordinate $x^{d+1}$. We use \gf\ for
defining the evolution  equation:
\eqn\gfss{\eqalign{&\partial _{d+1}g _{\mu \nu}=  R _{\mu \nu} -\demi g
_{\mu \nu} R+ \nabla  _{\{\m } A_{\n\}} +\kappa g _{\mu \nu} +H _{\mu
\nu} }}
  The fields now depend on $x^\mu$,
$1\leq \mu \leq d$, and $x^{d+1}$. 
Equation
\gfss\  can be seen from two different points of view.

The first point of view is that   this equation   describes  stochastic
quantization for the Einstein Lagrangian $\sqrt g(R+\kappa)$. $H _{\mu \nu}$ now has the heuristic interpretation    of a
stochastic noise. The  first two terms in the right hand side of   \gfss\ are nothing but
the Einstein equations of motion in $d$ dimensions. They  represent a drift force
along the physical degrees of freedom that balances the effect of the
noise $H _{\mu
\nu}$,  and determines   quantum gravity. Then,   the   term
$\nabla  _{\{\m } A_{\n\}}$ can be seen as a drift force along the gauge
orbits for reparametrization. Observables of gravity should not depend
on the choice of the vector field $A^\m$, which  is nevertheless
necessary for ensuring the convergence of the Langevin equation \gfss\
for all degrees of freedom in $ g _{\mu \nu}$.
 We can sum over all possibilities  for  
$A^\m$ which can be thus interpreted as the component  $ g ^{\mu ,d+1}$
of a metric in
$d+1$ dimensions. Then, \gfss\ should be read as:
\eqn\gfs{\eqalign{& \nabla_{ \{ d+1  } 
g _{\mu \nu \}   }  =  R _{\mu \nu}
-\demi g _{\mu \nu} R  +\k g _{\mu \nu} +H _{\mu \nu} }}

 This  generalizes the introduction of the additional component for the gauge field when one    quantizes   Yang--Mills theories with one
extra dimensions as in \zbgr. Moreover, we can introduce an arbitrary component 
$g^{d+1, d+1}$ as an unobservable     kernel that multiplies the right hand side 
of \gfs. Correlators in a slice cannot depend on $g^{d+1, d+1}$.
\gfs\ is formally a sort of Langevin equation that is covariant in 
$d+1$ dimensions. Once it is expressed in a supersymmetric
path integral,   it formally produces a TQFT based on the
action \gimp.
But now the interpretation is that $A^\m= g ^{\mu ,d+1}$ must be
gauged fixed as in \gfd\ to describe gravity in $d$ dimensions.   We will shortly get a much more attractive 
formulation for gravity in the next section, using the first order
formalism.

The second interpretation  is that \gfss\ is related to the 
renormalization group  equation with $x^{d+1}$ interpreted as a cut-off
or renormalization group parameter.  This is quite reminiscent   of
the ideas developed in \hverlinde.  One sees that for zero fluctuation,
$H_{\m\nu}=0$, the stochastic equations
\gfss\ reproduces the renormalization group equation used in \friedan.
Then, $g_{\mu,d+1}$ stands for the trivial  directions that are   represented by   reparametrizations of the metrics of the
$\sigma$-model.

\newsec{First order formalism and the link between gravity and
Yang--Mills theory}

We now give our argument that    $d$-gravity can be seen as  a
Yang--Mills theory in $d+1$ dimensions with $SO(d+1)$, $SO(d,1)$ or $SO(d-1,2)$ internal symmetry.
We   combine   the gravity first order formalism and  bulk
quantization.
\def\o{\omega} We   introduce the vielbein $e^a_\mu$ and spin
connection $\o^{ab}_\mu$.  The Lagrangian in $d$-dimensions is
\foot {We must replace 
$R^{ab}(\o)$ by $R^{ab}(\o)+\kappa e^a\wedge e^b$ for the case of gravity
with a cosmological constant.
}:
\eqn\einstein{\eqalign{L=\e^{ab a_3\ldots a_d} e^{a_3}\wedge
\ldots  e^{a_d} \wedge  R^{ab}(\omega) }} The gauge symmetry in the
indices $a,b,..$ is that  of the Poincar\'e group
$ISO(d)$ in the Euclidian case, or $ISO(d-1,1)$ in the Minkowski case. In both cases   
  $R^{ab}=d\o^{ab}+\o ^a_c\o^{cb}$ and $T^a=de^a+\o ^a_be^b$ are the
curvatures (Lorentz curvature and torsion).

Bulk quantization of   gravity  should be    based on the   
equations of motion  and the symmetries that one can infer from the gravity
action \einstein.  
 The equations of motion   determine   drift forces along transverse directions; they
must be supplemented by drift forces along gauge orbits for  local
Lorentz and translation invariance. The later are formally  equal to   to gauge
transformations, with   parameters that are replaced by arbitrary fields. We
call these new fields  $\o^{ab}_{d+1}$ and $e^{a}_{d+1}$. The reason
for this notation is that these fields  will be shortly   interpreted as
additional components for the spin connection and vielbein. This
generalizes \bautopr\ and \zbgr.  The bulk equations are thus:
\eqn\einstoc{\eqalign{
 \partial _{d+1} e_\mu^{a} =& ^*
\Big (
\e^{a  bc  a_4 \ldots a_d  }e^{a_4}\wedge
\ldots  e^{a_d} \wedge  R^{bc}  \Big )_\mu +D_\mu  e_{d+1}^{a} + \omega
_{d+1} ^{ab}  e^b_\mu +H^a_\mu
\cr
  \partial _{d+1} \omega _\mu^{ab} 
 =& ^*
\Big (
\e^{a  b  a_3 \ldots a_d  }e^{a_3}\wedge
\ldots  e^{a_d} \wedge  T^{b} \Big )_\mu +D_\mu  \omega^   {ab}_{d+1}
+H^{ab}_\mu }}
$H^{ab}_\mu$ and  $H^{a}_\mu$ can be heuristically understood as Gaussian
noises. However, we prefer to identify them as part of a BRST
topological quartet stemming from $\o^{ab} $ and  $e^{a} $, and
understand
\einstoc\ as the equation of motion with respect to $H$ of a TQFT
action. 
\einstoc\ can be rewritten as:
\eqn\einstoc{\eqalign{ T^a _ {[\mu, d+1]}=& ^*
\Big (
\e^{a  bc  a_4 \ldots a_d  }e^{a_4}\wedge
\ldots  e^{a_d} \wedge  R^{bc}  \Big )_\mu +H^a_\mu
\cr R^{ab}  _{[\mu, d+1]}=& ^*
\Big (
\e^{a  b  a_3 \ldots a_d  }e^{a_3}\wedge
\ldots  e^{a_d} \wedge  T^{b} \Big )_\mu +H^{ab}_\mu }}
$T^a$ and $R^{ab}$ are now two-forms in $(d+1)$-dimensional space that are valued in
$iso(d)$ or $iso(d-1,1)$. They lose their interpretations as  Poincar\'e  curvatures   in the $(d+1)$-dimensional bulk.

The formal implementation of  \einstoc\ in a path integral defines a
BRST invariant action in $(d+1)$-dimensional space of the topological type, with \einstoc\ as
topological gauge functions.   This  action is a sum of a pure
derivative along $x^{d+1}$ and a BRST-exact term. If we skip ghost of
ghost dependent terms, it is:
\eqn\einstop{\eqalign{
\int_{M_{d+1}} & {{\delta I}\over {\delta e^a}} D_{d+1}  e^a 
 +  {{\delta I}\over {\delta \o^{ab}  }}
D_{d+1} 
\o^{ab}
\cr & +  
\{ Q,
\bar\Psi^{a \mu} (T^a _ {[\mu, d+1]}- ^*
\Big (
\e^{a  bc  a_4 \ldots a_d  }e^{a_4}\wedge
\ldots  e^{a_d} \wedge  R^{bc}  \Big )_\mu +\demi  H^a_\mu \ )
\}
\cr & +  
\{ Q,
\bar\Psi^{a b\mu}
 R^{ab}  _{[\mu ,d+1]}- ^*
\Big (
\e^{a  b  a_3 \ldots a_d  }e^{a_3}\wedge
\ldots  e^{a_d} \wedge  T^{b} \Big )_\mu +\demi H^{ab}_\mu \ )
\} }} This is a  supersymmetric action. The important piece is the bosonic part,
since, for correlators that only depend  on the fields
$e$ and $\o$, the ghosts can be integrated out \zbgr. The relevant term
in 
\einstop\ is the bosonic part:
\eqn\einstopp{\eqalign{
\int_{M_{d+1}} dx^{d+1} dx^\mu  \  e\ \Big (
 T^a _ {\mu,{d+1}} T^{a  \mu ,   {d+1}} +
 T^a _ {\mu \nu} T^{a  \mu \nu}
 + R^{ab} _ {\mu,{d+1}} R^{ab  \mu,   {d+1}} +
 R^{ab} _ {\mu \nu} R^{a  b\mu \nu}\Big) }}
 $e=\det {e^a_\mu}$ and the $d$-dimensional indices are contracted by
the induced metrics 
$g_{\mu\nu}={e^a_\mu}{e^b_\nu}$.

We recognize the emergence of a Yang--Mills action in $d+1$ dimensions
with an  internal symmetry that is  $ISO(d)$ in the Euclidian case or
$ISO(d-1,1)$ in the Minkowski case. It is   natural to define a
metrics in $d+1$ dimensions
$G_{\a\b}$ that reads as $ g_{\mu\nu}=e^a_\mu e^a_\nu$,   $g_{\mu,d+1}=e^a_\mu e^a_{d+1}$ and $g_{d+1,d+1}=e^a_{d+1} e^a_{d+1} 
$. Then, defining $a$ as an $ISO(d)$  or $ISO(d-1,1)$ valued gauge
field in the $(d+1)$-dimensional space, and $F=da+a\wedge a$,  we can write 
\einstopp\ as: 
\eqn\einstopp{\eqalign{
\int_{M_{d+1}} d^{d+1} x \ \sqrt {G} \ 
 \tr \ (F_{\a\b} F^{\a\b}) }}
This action, supplemented by the supersymmetric terms, which    ensure
topological BRST symmetry, and which  can be  derived from
\einstop, formally determines  the correlation functions of gravity in $d$
dimensions. It must be inserted  
  in a path integral in $d+1$ dimensions, and one must choose  the observables in a given
slice, that is in a manifold of dimension $d$, suitably embedded in the
$d+1$-space. With the gauge  choice $g_{\mu,d+1}=0,g_{d+1,d+1}=1$,
the slice is at constant $x^{d+1}$.

We are almost at the point of finding natural   that   an action
  of the Yang--Mills type  with internal
$ SO(d+1)$, $ SO(d,1)$ or $ SO(d -1 ,2)$   invariance, with some sort of
supersymmetry,  is a good candidate for describing  gravity in $d$
dimensions.  We have indeed introduced  an additional world coordinate $x^{d+1}$;
then, the   internal Yang--Mills invariance   $ SO(d+1)$, $ SO(d,1)$ or $ SO(d-1, 2 )$   can  now  be postulated as the internal  symmetry of \einstopp, which we would like to consider as a starting point. This 
  necessitates   accepting    that bulk quantization is a defining principle.
 The 
$ISO(d)$  or $ISO(d-1,1)$   symmetries that occur in euclidian or    minkowskian gravity should arise as   particular effects  of the quantum field theory  that one builds from   \einstopp. Non trivial renormalizations could provide the relevant   contractions    of $ SO(d+1)$, $ SO(d,1)$ or $ SO(d -1 ,2)$. 
 
\newsec{Discussion}

We   have suggested  an intriguing relation between gravity in $d$ dimensions and
   topological Yang--Mills theory with internal $SO(d+1)$, $SO(d,1)$ or
$SO(d-1,2)$ invariance in $d+1$ dimensions. The internal bulk 
coordinate  is   non compact and  can be heuristically understood as the 
time  of stochastic 
 quantization. The   observables  of ordinary gravity can live in any
given d-slice of the bulk,   not only in its    boundary. This
suggests  iterating  the construction, by considering  
sub-slices, in which gravity  would descend from  another theory. Having theories that are related by jumps of two units could be of interest for new interpretations of descent equations for anomalies.

Our observation  is
compatible with the description of 3D-gravity by a Chern--Simons action
\Witgravcs, since  the latter    can also be directly formulated as the
topological Yang--Mills action in four dimensions with $SO(4)$ internal
symmetry 
 \baugrav. 

In the general case,
$D\neq 3$,   a contraction in the Lie algebra    is required   
   for  recovering  the 
Poincar\'e symmetry. This implies  a symmetry breaking.  Non perturbative effects in the bulk could   trigger
  singular renormalization effects      that   change  the structure coefficients and determine eventually the contractions.  
 Bulk quantization, which is a generalization of stochastic quantization, does not require that the non compact extra dimension be flat. A natural parameter is the curvature $R$ along this direction, and the contractions that we invoke might appear in the limit  $R^{-1}\to \infty$.

 \vskip .5cm
{\centerline{\bf Acknowledgments}}

It is a pleasure to thank's my collaborator Daniel Zwanziger with whom I have  discussed many ideas concerning quantization with a bulk time, as well as the  members of the    Enrico Fermi Institute for their hospitality and very interesting
discussions  concerning this work.  
  \vskip .5cm


\footatend\vfill\supereject\immediate\closeout\rfile\writestoppt
\baselineskip=14pt\centerline{{\bf References}}\bigskip{\frenchspacing%
\parindent=20pt\escapechar=` \input refs.tmp\vfill\eject}\nonfrenchspacing

\bye